\documentclass[preprint,12pt]{elsarticle}
\usepackage[utf8]{inputenc}

\usepackage{amssymb}
\usepackage{graphics}
\usepackage{graphicx}
\usepackage{amsmath}

\usepackage{acronym}
\usepackage{subfig}
\usepackage{tabularx}
\usepackage{url}

\journal{-}

\begin{document}

\begin{frontmatter}



\title{Geometric Design of Micro Scale Volumetric Receiver Using System-Level Inputs: An Application of Surrogate-Based Approach}



\author[Ozu,CEEE]{Tufan Akba\corref{cor1}}\ead{tufan.akba@ozu.edu.tr}
\cortext[cor1]{Corresponding author.}
\author[Metu,GUNAM]{Derek K. Baker}
\author[Ozu,CEEE]{M. Pınar Mengüç}

\affiliation[Ozu]{organization={Department of Mechanical Engineering},
            addressline={Ozyegin University}, 
            city={Istanbul},
            country={Turkey}}

\affiliation[CEEE]{organization={Center for Energy, Environment and Economy (CEEE)},
            addressline={Ozyegin University}, 
            city={Istanbul},
            country={Turkey}}

\affiliation[Metu]{organization={Department of Mechanical Engineering},
            addressline={Middle East Technical University}, 
            city={Ankara},
            country={Turkey}}

\affiliation[GUNAM]{organization={Center for Solar Energy Research and Application (GUNAM)},
            addressline={Middle East Technical University}, 
            city={Ankara},
            country={Turkey}}

\begin{abstract}
Concentrating solar thermal power is an emerging renewable technology with accessible storage options to generate electricity when required. Central receiver systems or solar towers have the highest commercial potential in large-scale power plants because of reaching the highest temperature. With the increasing solar chemistry applications and new solar thermal power plants, various receiver designs require in micro or macro-scale, in materials, and temperature limits. The purpose of the article is computing the geometry of the receiver in various conditions and provide information during the conceptual design. This paper proposes a surrogate-based design optimization for a micro-scale volumetric receiver model in the literature. The study includes creating training data using the Latin Hypercube method, training five different surrogate models, surrogate model validation, selection procedure, and surrogate-based design optimization. Selected surrogates have over $98\%$ R\textsuperscript{2} fit and less than $4\%$ root mean square error. In final step, optimization performance compared with the base model. Because of the model complexity, surrogate models reached better objective values in a significantly shorter time.
\end{abstract}




\end{frontmatter}


\section*{List of Symbols}
\begin{acronym}[MMMMM] 

\acro{$c_p$}{specific heat at constant pressure}{ $[J/kg \cdot K]$}
\acro{$h$}{heat transfer coefficient}{ $[W/m^2 \cdot K]$}
\acro{$k$}{thermal conductivity}{ $[W/m \cdot K]$}
\acro{$q''$}{heat flux}{ $[W/m^2]$}
\acro{$r$}{inner radius}{ $[m]$}
\acro{$L$}{length}{ $[m]$}
\acro{$s$}{specific surface}{ $[m^{-1}]$}
\acro{$T$}{temperature}{ $[K]$}
\acro{$v$}{volume}{ [$m^3$]}
\acro{rho}[\ensuremath{\rho}]{density}{ $[kg/m^3]$}


\end{acronym}
\section*{Subscripts}
\begin{acronym}[MMMMM] 


\acro{$f$}{fluid}


\acro{$s$}{solid}
\acro{$r$}{radiation}

\end{acronym}
\section*{Acronyms}
\begin{acronym}[MMMMM]

\acro{CST}{concentrating solar thermal}
\acro{DOE}{design of experiments}
\acro{HCE}{heat collecting elements}
\acro{HTF}{heat transfer fluid}
\acro{LHD}{Latin hypercube design}
\acro{MDO}{multidisciplinary design optimization}
\acro{RBF}{radial basis function}
\acro{RMSE}{root mean square error}
\acro{SLSQP}{sequential least squares programming}
\acro{TES}{thermal energy storage}

\end{acronym}

\section{Introduction}
\label{sec:intro}
Thermodynamic cycle analysis gives insight into the overall system performance and shows irreversibilities in each process creating the system \cite{DEMIREL2007275}. For a system design (i.e. power plant, gas turbine, internal combustion engine), thermodynamic cycle analysis is the top-level model and defines the expected performance of the components as the initial phase. After the component design is completed, the cycle analysis is performed with the new values to observe the system-level impact for providing feedback information. Thermodynamic cycle analysis is the core calculation for systems in different performance parameters, design limitations, and operation conditions. Fossil to renewable transition changed the analysis structure and performance parameters, design limitations, and operation conditions. The known method from the fossil-fueled systems, fuel is the controlled input for optimization of the plant efficiency by adjusting the fuel \cite{Oates1985,DIGIANFRANCESCO20171,heywood}. Unlike fossil fuels, renewable energy sources are not controlled inputs. The thermodynamic models focus on maximizing cumulative power generation by creating multiple operation points instead of optimizing the most efficient configuration \cite{MAKI2012101, AKBA2020497}.

For \ac{CST} power, solar field, auxiliary heater, \ac{TES}, and power block are the main components of the plant. The possible combinations of these components define the multiple operation points of the thermodynamic analysis \cite{WAGNER20141652}. Defining design requirements gets more complex in multiple operation points. System-level simulations are required for verification of the component design at multiple points. The current study focuses on component design integration in the thermodynamic cycle analysis. The coupled design (thermodynamic system performance and integrated component design) will solve the problem and satisfy these requirements together. However, coupled modeling requires \ac{MDO} and requires a problem architecture rather than solving every step separately \cite{MDAOArchitecture2013}.

The need for \ac{MDO} arose to solve multiple subproblems in different disciplines for complex engineering systems. \ac{MDO} helps to solve these systems in two ways. The first is coupling the system with all the interdisciplinary interactions. The second way is optimizing all the design variables coupled, and the trade-offs of the subproblem design considerations \cite{OpenMDAO2019}. The optimization problem may diverge because of the computational load while solving the subproblems simultaneously or due to the structure of the optimization problem. One alternative to converge the optimization problem is using an approximate (called surrogate) model that retains sufficient accuracy to represent model complexity in an error range. Surrogate modeling has advantages for gradient-based optimization, especially when the model has a high variation of gradients (i.e. noisy data) \cite{martins_ning_2021}.

Surrogate models or metamodels are used for simplifying complex engineering models. These models are less accurate but can provide a fast alternative to original models. The accuracy of the model is estimated before using the model. A surrogate model is evaluated by its computational time and accuracy \cite{Alizadeh2020}. Several surrogate models can be found in the literature. Response surface methodology \cite{BoxGE2005} is one the fastest method for surrogate modeling and kriging \cite{Kriging} is the most common alternative originating from mining applications. Because of the computational load of the kriging, first or second order response surfaces created to observe the surrogate performance and accuracy of the surrogate model increases using kriging in most of the cases. Neural networks are another surrogate model method that provides solution alternatives in the form of chains of simple functions. These chains are called networks, and each calculation node is called a neuron \cite{IBMNeuralNetworks}. The accuracy of the surrogate model is highly dependent on the sample or training data. The term "\ac{DOE}" focuses on reflecting the model behavior by screening the required number of samples called experiments \cite{4839625}. There are several sampling methods like \ac{LHD}, factorial designs, random selection, orthogonal arrays and low-discrepancy sequences \cite{Corchado2007,Montgomery2019,Alizadeh2020, martins_ning_2021} for training accurate surrogate models.

One advantage of surrogate modeling is the efficient solution of multifidelity problems. It maintains the required model complexity in the design phase, and the surrogate model transfers the required information for system optimization. The effectiveness of this approach was observed in several studies in different designs such as battery thermal management system \cite{WANG2021102771}, permanent magnet synchronous motor \cite{LI2021101203}, battery package optimization \cite{XU2021121380}. A similar approach is the core idea explained in the article. The actual model is used for an accurate solution. 
Training data is generated using the model for surrogate training and optimization is performed using the surrogate model.

The current study explains an integrated way of receiver modeling. Receivers are the solar radiation collecting elements in \ac{CST} power plants. Several receiver models are existing in the literature \cite{Hischier2012, GODINI2021574, CAPUANO2016656} and ETH's receiver model \cite{eth_rec} is replicated in this study. The main purpose of the article is to fill the gap in component design in \ac{CST} technology, especially in multiple operating conditions, and represent the results for proving that surrogate modeling is an efficient way of finding the optimum solution in complex design problems. In the article, different surrogate models are trained and their performances are compared with the base model (replicated model used cases) for validation purposes. After valid surrogates are selected, optimizations are performed with the selected surrogate model and the base model. Optimization results and the final optimum points are compared as the output of the study. The content of the article is structured as follows: In section \ref{sec:statement}, governing equations of the replicated receiver model and problem statement are explained. Section \ref{sec:design} focuses on building the surrogate model and optimization. Performance metrics, other findings, results, and discussion are in Section \ref{sec:results}. The final remarks and conclusion are in Section \ref{sec:conc}.

\section{Problem Statement and Numerical Model}
\label{sec:statement}
Receivers have three distinct designs: In external receivers, liquid \ac{HTF} passes through the \ac{HCE} which are exposed to concentrated solar radiation. In solar tower (or central receiver system), external receivers are widely used in large-scale power generation from Solar One, the pilot central receivers system in the late 1970s, to recent state-of-the-art large-scale plants including Gemasolar, Crescent Dunes, Noor3, Delingha, and others \cite{ASSELINEAU2022354}. Internal receivers are enclosed, and solar radiation is concentrated in an opening. These receivers are selected in solar tower power plants if the reflector field is directional. \ac{HCE} are enclosed for decreasing the ambient losses \cite{DINCER2014369}.

The volumetric receivers are another receiver preferred for gaseous \ac{HTF} applications. These receivers use porous structures for efficient heat transfer and unlike other receivers, they do not absorb the solar radiation outside of a tube or a surface \cite{Romero2002}. For hydrogen generation and other solar chemistry applications, volumetric receivers are used as a reactor of the plant \cite{ASSELINEAU2022354}. Focus of the study is designing a gaseous (air) fluid passing volumetric receiver in micro scale, which operates around $1kW$ concentrating solar radiation.

The objective of this study is to model a volumetric receiver using the system-level inputs. For receiver design, \ac{HTF} inlet conditions are critical information for efficient receiver design and resultant final plant design. Since the flow parameters are determined by the system's other components such as the compression ratio or inlet conditions, optimal receiver design changes with changing inlet conditions. For demonstration purposes, one of the volumetric receiver models is selected from the literature \cite{eth_rec}, and the performance is improved using surrogate model optimization.

\subsection{Receiver Model}
\label{ssec:receiver}

\begin{figure}[htbp]
    \centering
    \includegraphics[width=0.7\textwidth]{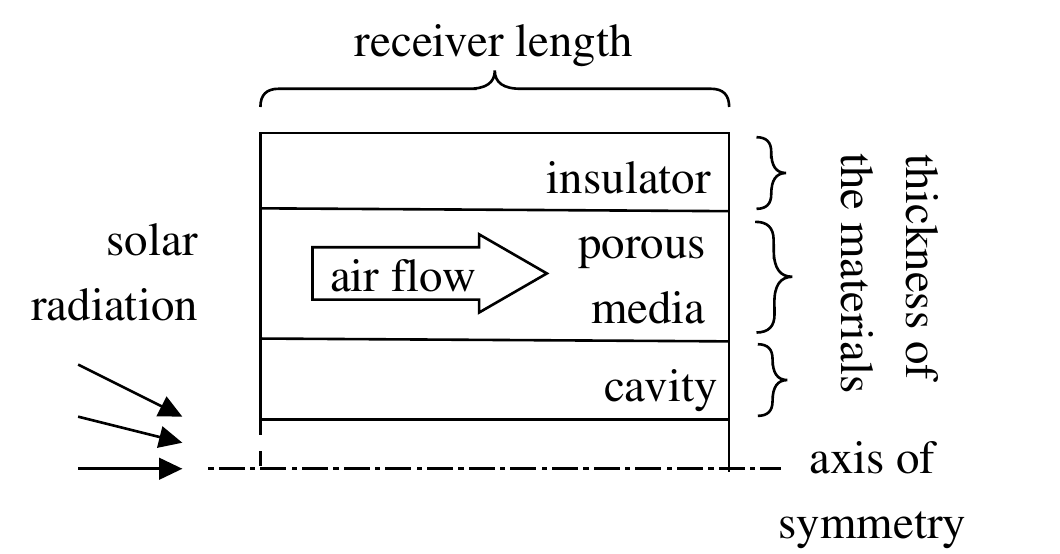}
    \caption{Asymmetric volumetric receiver model used in the article}
    \label{fig:model}
\end{figure}

Receiver model illustration is shown in Figure \ref{fig:model}. Concentrated solar radiation focuses on the cavity center of the receiver. Radiation is absorbed and conducted through the cavity. Air as \ac{HTF} is passing through the porous media. The outer layer of the porous media is insulated for decreasing convective losses.

Receiver model solves steady-state energy balance equations given in Equation \ref{for:solid},\ref{for:fluid}. Details of the model and further information are given in Reference \cite{eth_rec}. The model can be summarized as follows: Solar radiation hits the inner surface of the cavity. Using radiosity analysis, the heat fluxes of the inner surface are calculated iteratively. Heat is conducted through cavity material and in porous media heat exchanges between solid and fluid material. Inside the porous media, combined conduction, convection, and radiation heat transfer modes are solved. At the outer layer of the cylinder, heat is conducted through the insulator. Convective heat loss dissipated from the outer surfaces as the rest of the boundary conditions.

\begin{flalign}\label{for:solid}
\frac{1}{r}\frac{\partial}{\partial r}\left(rk_s\frac{\partial T}{\partial r}\right)+ \frac{\partial}{\partial z}\left(k_s\frac{\partial T}{\partial z}\right)-sh(T_s-T_f)+q_r'''=0
\end{flalign}

\begin{flalign}\label{for:fluid}
\rho v c_p\frac{\partial T_f}{\partial z}=sh(T_s-T_f)
\end{flalign}

In Equation \ref{for:solid} and \ref{for:fluid}, subscripts $s$, $f$ refers to solid and fluid parts. Solid parts can be cavity, porous media and insulator. Fluid (air) is only passing through the porous media and fluid solution (Equation \ref{for:fluid} is only solved for porous media region. The equation is axisymmetric model and solved simultaneously in radial direction $r$ and axial direction $z$. Convective heat transfer mode is defined in both equations. The terms $s$ and $h$ are specific surface and heat transfer coefficients, radiative heat flux defined in $q'''_r$ and in this equation $P_1$ method is used \cite{menguc2020}. In Equation \ref{for:fluid}, $\rho$, $\nu$ and $c_p$ are density, volume and constant specific heat, respectively.

In the scope of the article, the receiver model is used for accurate solutions in the design condition and for creating training data. In the solution step, the receiver model is used for steady-state temperature distribution throughout the domain. In design optimization, the surrogate model is used for parameter estimation for given conditions and constraints.

\subsection{Problem Statement}
\label{ssec:statement}

Receiver design depends on several conceptual aspects: It is the component reaching the maximum temperature in a \ac{CST} plant. It is directly correlated with Carnot efficiency and increasing the temperature is a key factor in increasing the overall thermal efficiency. However, the high-temperature nature of the receiver suffers high radiative losses which decreases the efficiency. Therefore, there is an optimum temperature for an ideal receiver depending on the concentrated solar radiation \cite{STEINFELD2005603}. In solar chemistry, reactions have different temperature and pressure ranges, which defines the receiver design requirements. As an example, for solar hydrogen reforming, receiver temperature and pressures can be significantly different, and the final design may have separate receivers (i.e. for reduction step at 1500\textsuperscript{0}C and 0.1 mbar and oxidation step at 900\textsuperscript{0}C and 1 mbar) \cite{Schappi2022}. Another aspect is the material limitations, which have to operate within the allowable temperature limits.

\begin{figure}[htbp]
    \centering
    \includegraphics[width=0.7\textwidth]{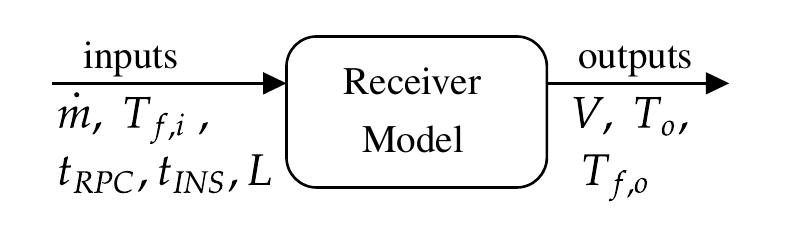}
    \caption{Receiver model inputs and outputs for data generation}
    \label{fig:surrogate}
\end{figure}

The problem is designing the receiver in desired conditions by using the receiver model explained in the previous section. For surrogate modeling, required data (training and validation) is generated by solving the receiver model which calculates several outputs as described in Figure \ref{fig:surrogate} for given inputs. In the figure, $\Dot{m}$ is the air mass flow rate, and inlet and outlet air temperatures are $T_{f,i}$ and $T_{f,o}$, respectively. Design parameters such as porous media and insulation thicknesses are $t_{RPC}$ and $t_{INS}$. $L$ and $V$ are the length and volume of the receiver. For decreasing the convective losses outer surface temperature of the receiver is limited and it is denoted as $T_o$.

As an optimization problem, problem statement is formulated as shown in Equation \ref{for:opt}. For physical limitations, every geometric variables have upper and lower limits denoted as underline and overline. Overall dimensions are constrained by maximum volume of the receiver.

\begin{equation}
\label{for:opt}
\begin{split}
\text{maximize} & \quad T_{f,o} \\
\text{by varying} &  \quad \underline{t_{RPC}} \leq t_{RPC} \leq \overline{t_{RPC}}\\
& \quad \underline{t_{INS}} \leq t_{INS} \leq \overline{t_{INS}}\\
& \quad \underline{L} \leq L \leq \overline{L}\\
\text{subject to} &\quad V - V_{max} \leq 0 \\
 &\quad T_{o} - T_{o,max}\leq 0\\
 \end{split}
\end{equation}

For a micro receiver design in 1000W concentrated solar radiation input, problem parameters are listed in Table \ref{tab:inputs}.

\begin{table}[!h]
\centering
\begin{tabular}{ccc}
\hline 
 Parameter & Unit & Limit \\
\hline 
$\displaystyle \dot{m}$ & $\displaystyle kg/s$ & $\displaystyle 0.00066$ \\
$\displaystyle T_{f,i}$ & $\displaystyle K$ & $\displaystyle 300$ \\
$\displaystyle t_{RPC}$ & $\displaystyle m$ & $\displaystyle 0.005\ -\ 0.025$ \\
$\displaystyle t_{INS}$ & $\displaystyle m$ & $\displaystyle 0.05\ -\ 0.25$ \\
$\displaystyle L$ & $\displaystyle m$ & $\displaystyle 0.02\ -\ 0.1$ \\
$\displaystyle V_{max}$ & $\displaystyle m^{3}$ & $\displaystyle 0.00375$ \\
$\displaystyle T_{o,max}$ & $\displaystyle ^o C$ & $\displaystyle 100$ \\
\hline
\end{tabular}
    \caption{Defined values and limits for the problem statement}
    \label{tab:inputs}
\end{table} 

\section{Design Methodology and Optimization}
\label{sec:design}
Design optimization of complex engineering problems in multiple variables significantly increases the number of solutions and takes high computational time. However, for a sample of solutions, variable trends can be approximated. These approximate functions (called surrogate models) significantly speed up the optimization and can be easily validated by implementing to real problem \cite{CROMBECQ2011683, BATMAZ2003455}.

\subsection{Creating Training Data}
\begin{figure}[htbp]
    \centering
    \includegraphics[width=1.0\textwidth]{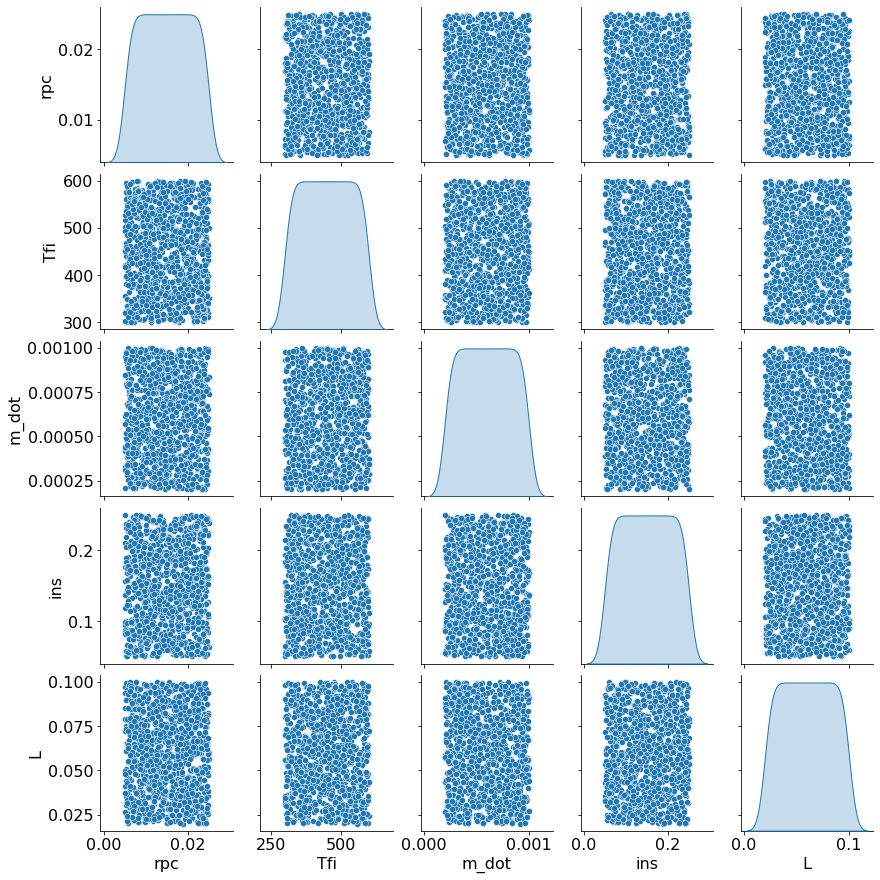}
    \caption{Uniform distribution of design variables (\ac{DOE} inputs with 1024 samples)}
    \label{fig:training_inputs}
\end{figure}

\begin{figure}[htbp]
    \centering
    \includegraphics[width=0.6\textwidth]{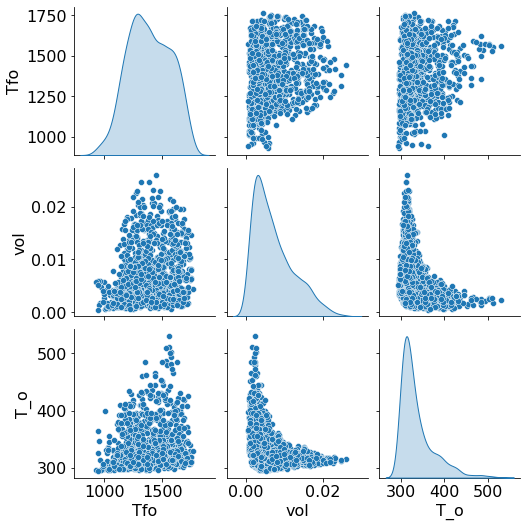}
    \caption{Distribution of \ac{DOE} results}
    \label{fig:training_outputs}
\end{figure}

Surrogate models are created from the training data. Space filling of the design space is the primary consideration of modern \ac{DOE} methods. In current study, \ac{LHD} \cite{McKay1979} is selected. This method is implemented in the PyDoe2 package in Python and wrapped in OpenMDAO - DOEDriver \cite{OpenMDAO2019}. For the given $5$ inputs shown in Figure \ref{fig:surrogate}, $4^5$ (1024) training data is created in design space. Distribution of the input data is shown in Figure \ref{fig:training_inputs} and \ac{DOE} results are shown in \ref{fig:training_outputs}. The training data is the combination of these two figures.

\subsection{Surrogate Modeling}

Surrogate model accuracy should be validated before use. Validation gives an idea about the error margin of the surrogate and it prevents overfitting problems. In this study, training data did not split as training-validation sets or performed k-fold cross-validation \cite{Optimization}. Rather, a new \ac{DOE} is performed with $100$ points as a validation set.

Second-order response surface, nearest neighbor (N-dimensional interpolation), and kriging algorithms are used for developing surrogate models. Three different interpolations (linear, weighted, \ac{RBF}) are used in the nearest neighbor algorithm. Overall, five surrogate models were created. They are validated using prediction error histogram plots. True values are plotted with predictions for observing the error distribution. In all cases, the same training and validation sets are used for benchmarking the surrogate model performance.

\subsection{Optimization}

As described in Section \ref{ssec:statement}, optimization problem has three geometric variables and volume constraint as defined in Equation \ref{for:opt}. Inlet conditions (mass flow rate and temperature) are fixed at the start of the optimization case, which is defined by other components (i.e. compressor) of the plant. Surrogate models are considered black-box models in optimizations. Among different open-source optimizers in SciPy \citep{SciPy}, an open-source Python library used for scientific and technical computing, \ac{SLSQP} algorithm is selected. \ac{SLSQP} is a gradient-based, bounded, constrained local optimization algorithm \citep{SLSQP}. Because of the black-box modeling of surrogates, the optimizer calculates the gradients. In the final step, surrogate-based and base mode optimizations are compared using the same optimizer (\ac{SLSQP}) and starting with the same initial values.

\section{Results and Discussion}
\label{sec:results}
In this section, the results are shown in the following order. First, the results of the different surrogate model algorithms are shown. After the surrogate models are selected, optimizations are performed with the selected surrogates and the base model. Surrogate results and optimizations are discussed in performance and accuracy aspects.

\subsection{Surrogate Model Results}

\begin{figure*}[htbp]
   \centering
\begin{tabular}{cc}
\subfloat[Histogram of the prediction error using kriging method]{\includegraphics[width=6.5cm]{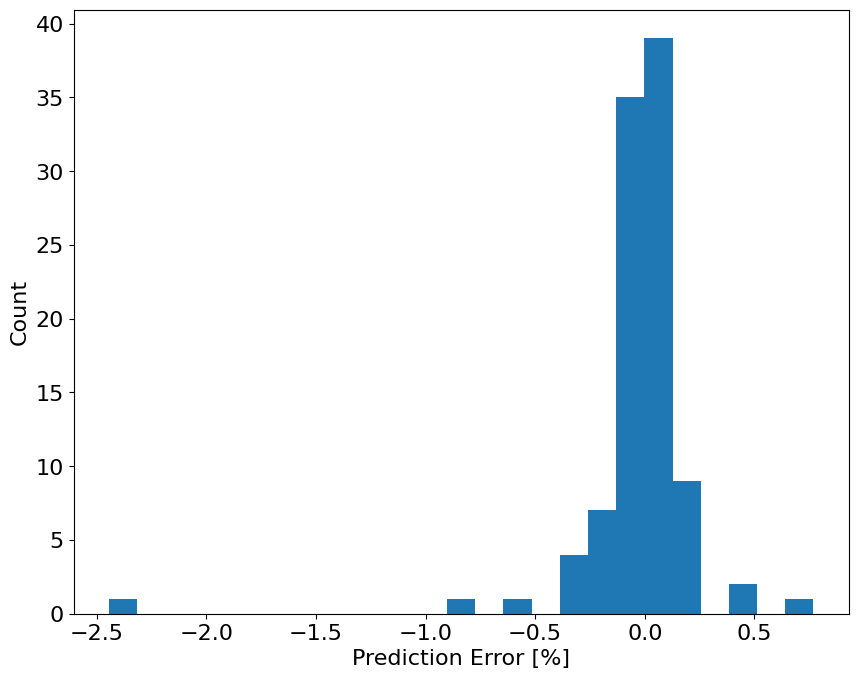}} &
\subfloat[Predictions of the validation data using kriging method]{\includegraphics[width=5.5cm]{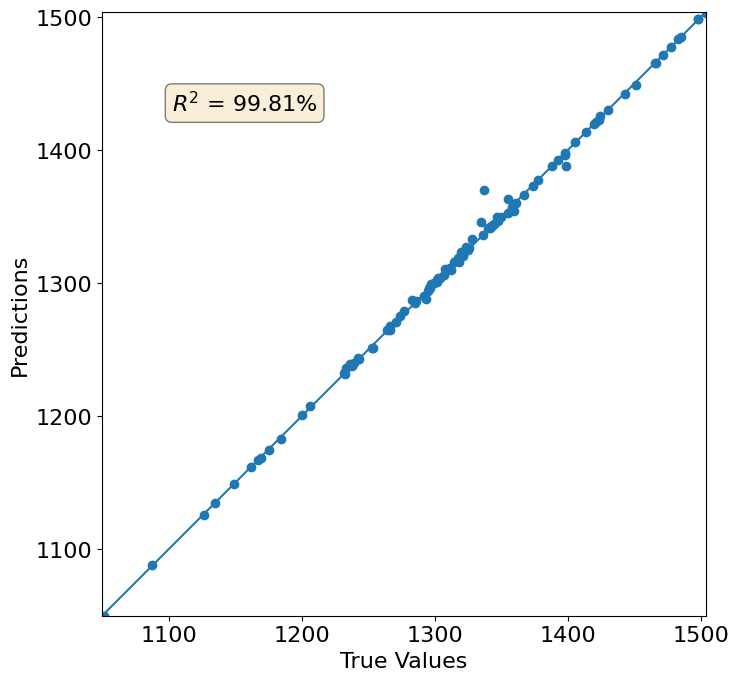}} \\
 \subfloat[Histogram of the prediction error using response surface method]{\includegraphics[width=6.5cm]{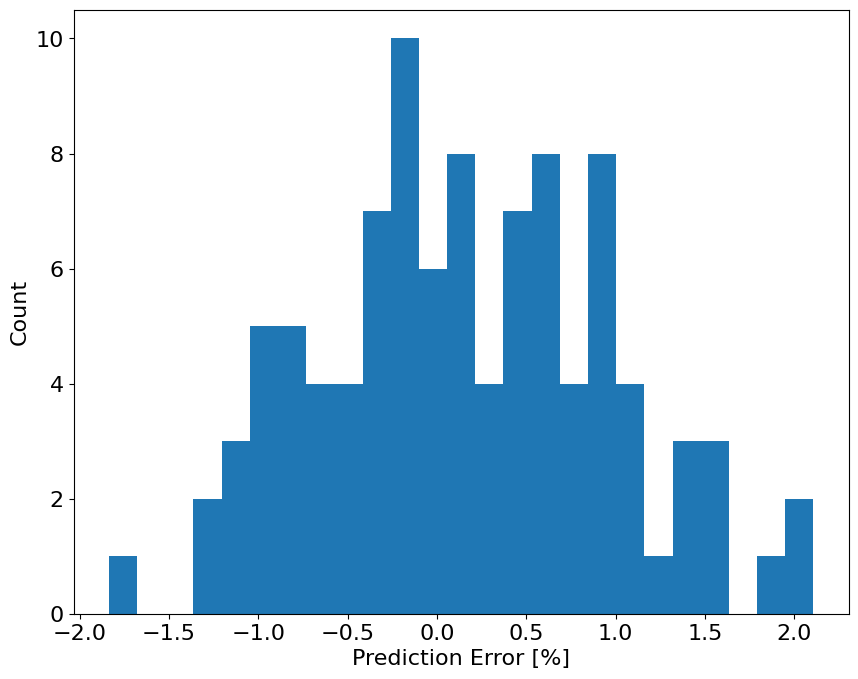}} &
\subfloat[Predictions of the validation data using response surface method]{\includegraphics[width=5.5cm]{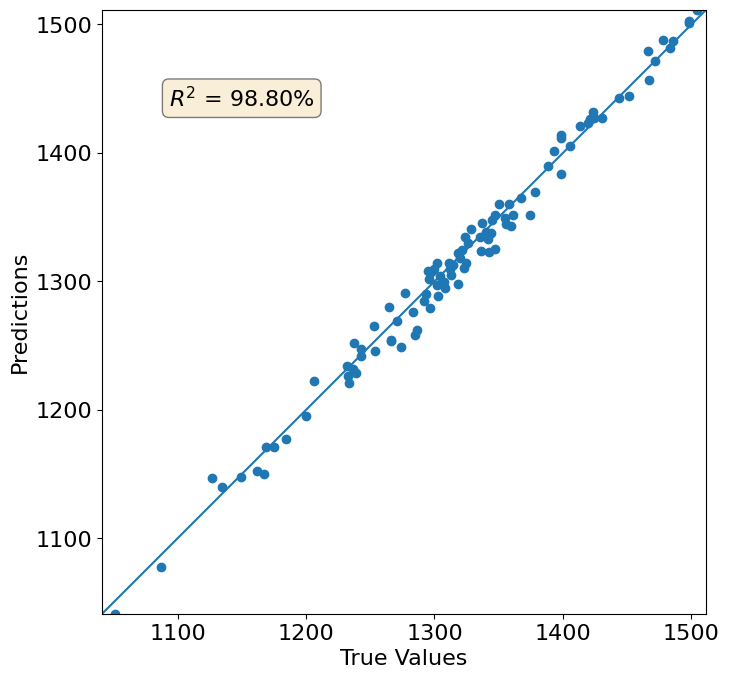}} \\
\end{tabular}
   \caption{Validation results of kriging and response surface surrogate models}
   \label{fig:results_RS_krig}
\end{figure*}

Surrogate models are compared with validation data. In this section, kriging and response surface models are shown in Figure \ref{fig:results_RS_krig} and nearest neighbor results are shown in Figure \ref{fig:results_NN}. Nearest neighbor and response surface models are significantly faster (over 100 times) than kriging. 

The response surface result is in $\pm2\%$ error range  and $98.8\%$ R\textsuperscript{2} fit with validation data and error margin decreases to $\pm1\%$ and $99.8\%$ fit with less than $4\%$ \ac{RMSE} in kriging model. Nearest neighbor results have similar calculation time with response surface results but worse accuracy. Validation results of the surrogate model is shown in Table \ref{tab:surr_res}.

\begin{table}[htbp]
\centering
\begin{tabular}{lcccccc}
\hline 
  &  & Response & & \multicolumn{3}{c}{Nearest Neighbor (fits)}\\
  & Units & Surface & Kriging &  Linear & \ac{RBF} & weighted\\
\hline 
 R\textsuperscript{2} &  \%  &   98.80  &   99.81 &  28.34 & 0.77 & 87.32 \\
 \ac{RMSE} &  K  &  10.7 &  4.0 &   147.9 &  378.7 & 34.8 \\
\hline
\end{tabular}
\caption{Validation results of the models}
\label{tab:surr_res}
\end{table}

Error margin has high variation in nearest neighbor method. While predictions are close throughout the validation set in kriging and response surface. The trend of the predictions is significantly different in each nearest neighbor method. Because of high variation, for a quick estimate of surrogate model performance, response surface models are advised. Surrogate model performance can be improved in two ways. The first method is increasing the training data samples. This approach has limitations because of using a second-order surrogate model. As modeling is replicated from 1024 training samples to 10000 samples and similar results ($\pm2\%$) are observed in the error margin. The second method is changing the surrogate model to kriging, which requires more time to build the surrogate model. Kriging decreases the prediction error. The first method applied to kriging and cannot build a surrogate model. Therefore, the user is advised to start building kriging models with relatively small training data.

From the result of the surrogate models, kriging and response surface models are selected for solving the optimization problem.

\subsection{Design Optimization Results}

Gradient-based optimizers are sensitive to gradient variation among the input and output parameters. These optimization problems are solved by scaling the gradients. As a rule of thumb, it is advised to arrange the gradients to similar ranges. Because of the model complexity, even the optimizer tuned for scaling at the beginning; during the solution, scaling may have an adverse effect to diverge the optimization. On the other hand, surrogate models are simplified and less tend to gradient oscillations. Fewer gradient oscillations may prevent diverging or lead to a faster and better performing optimization. In some cases, surrogate models converged better points. In these cases, we observed that, in the same initial values, base model optimization has bad scaling during the optimization. In some of those values, the surrogate model has relatively better scaling performance.

\begin{table}[!h]
\centering
\begin{tabular}{lcccc}
\hline 
  &  & Response & & Base \\
  & Units & Surface & Kriging &  Model\\
\hline 
$ t_{RPC} $ & $  m$ & $  0.0086$ & $  0.0135$ & $ 0.0109 $\\
$ t_{INS} $ & $  m$ & $  0.0995$ & $  0.0853$ & $ 0.0786 $\\
$ L$ & $  m$ & $  0.0729$ & $  0.0807$ & $ 0.0630 $\\
$ \mathbf{T_{f,o}}$ & $  K$ & $  \mathbf{1324.7}$ & $  \mathbf{1383.9}$ & $\mathbf{1292.7} $\\
\hline
\end{tabular}
\caption{Optimization results of the models}
\label{tab:opt_res}
\end{table}

Optimization results are listed in Table \ref{tab:opt_res}. Surrogate-based optimizations are performed using kriging and response surface algorithms. Building a kriging surrogate model takes significant time. However, created surrogates have similar iteration times during the optimizations. The base model is not a surrogate, it is using receiver model with \ac{SLSQP} optimizer for reference. Outlet fluid temperature is the maximized objective by varying other values in the table.

\section{Conclusion}
\label{sec:conc}
This paper investigates an alternative way of receiver design optimization. A receiver design from the literature is reproduced using OpenMDAO framework. Using Latin hypercube method, two different number of training set and one validation set is created. Using those sets five surrogate models are created and validated. Kriging gave the most accurate result, and second-order response surface provided very accurate results less than $1\%$ of the computational time. Design optimizations performed using these two surrogates and base (no surrogate) cases. Findings of the article is listed below:

\begin{enumerate}

    \item\textbf{Surrogate models:}
    Kriging and response surface model have less than $2\%$ prediction error. By using response surface model, a fast and accurate surrogate model can be build and tested in a short time. This model can be improved by changing the surrogate algorithm to kriging. Building kriging model takes relatively more time (in different number of training data, time difference between response surface to kriging about 100 times), and model may not build after a sampling limit. When performance risks and computational load are considered, nearest neighbor algorithm is not suggested. Different fit options changes the model behavior drastically even same datasets are used.
    
    \item \textbf{Optimization - Sampling:}
    Simplification due to the surrogates decreases the computational load and turns the model used in optimization into a black-box model. Surrogates allow faster iterations and very low computational load especially in gradient calculation steps. However, optimization may converge to a low-sampled design space and error may be higher than the validation step. Even the surrogates are validated, it is suggested to solve optimal values in receiver model.
    
    \item \textbf{Optimization - Scaling:}
    As shown in Table \ref{tab:opt_res}, optimization problems are scaled and solved with the same initial values. For a complex model, scaling becomes hard to control and optimization required more iterations to converge. When the calculated gradients are checked, surrogate models have consistent gradients and base model does not have during the solution.

    \item \textbf{Optimization - Results:}
    As seen in Table \ref{tab:opt_res}, surrogate based optimization designed better receiver for given receiver model. When time advantage of the surrogate modeling is considered, explained solution throughout the article provides better results in less time.
\end{enumerate}


\section*{Declaration of Interest}
The authors declare that they have no known competing financial interests or personal relationships that could have appeared to influence the work reported in this paper.

\section*{Acknowledgement}
For replication purposes, source code is uploaded on \url{https://github.com/TufanAkba/surrogate_article.git}.

\appendix

\section{Nearest Neighbor Surrogate Model Results}
\label{sec:appendix}
The results shown in this section has not accurate results. The reason of adding those results to the study is showing the variation of the surrogates by changing the interpolant type and demonstrating the variation of the data shift and prediction error distribution.

\begin{figure*}[htbp]
   \centering
\begin{tabular}{cc}
\subfloat[Histogram of the prediction error using linear interpolant]{\includegraphics[width=6.0cm]{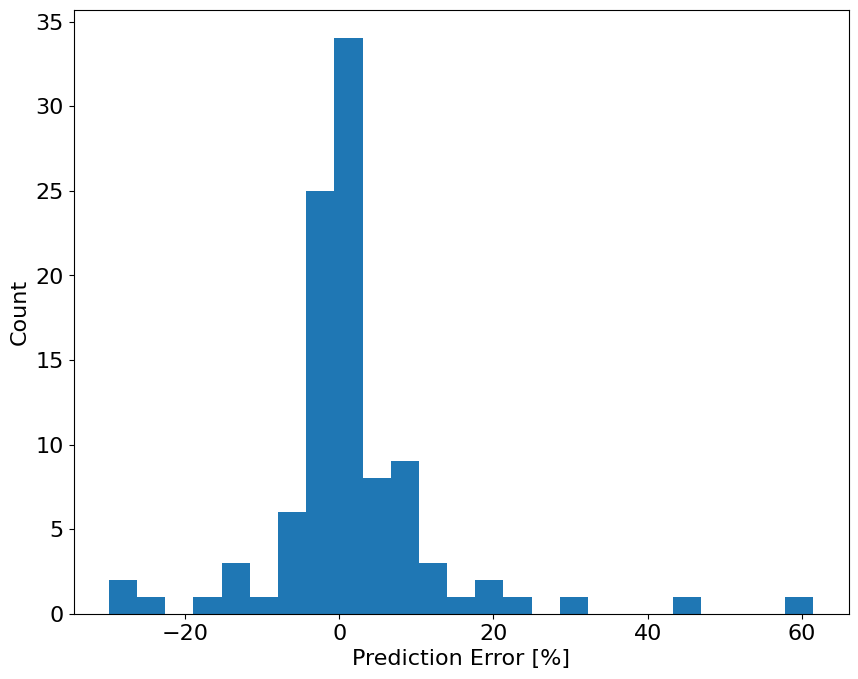}} &
\subfloat[Predictions of the validation data using linear interpolant]{\includegraphics[width=5.0cm]{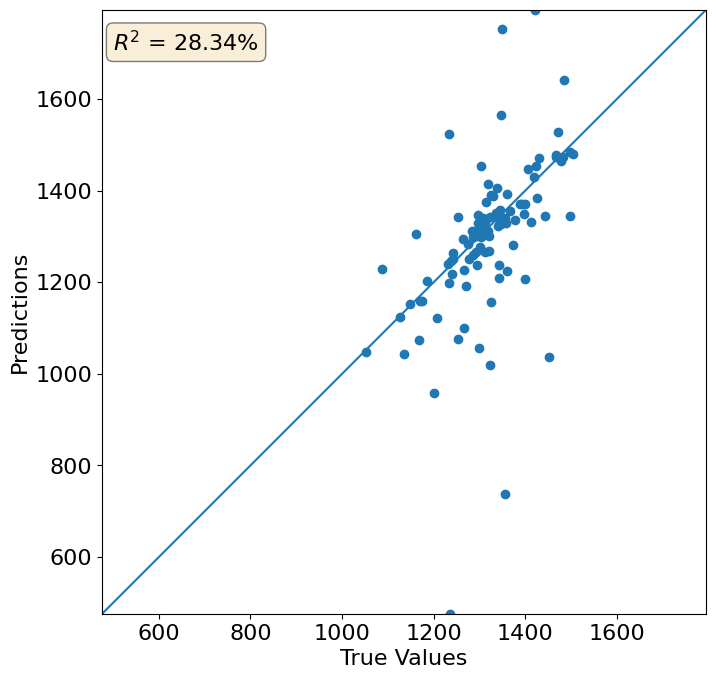}} \\
\subfloat[Histogram of the prediction error using \ac{RBF} interpolant]{\includegraphics[width=6.0cm]{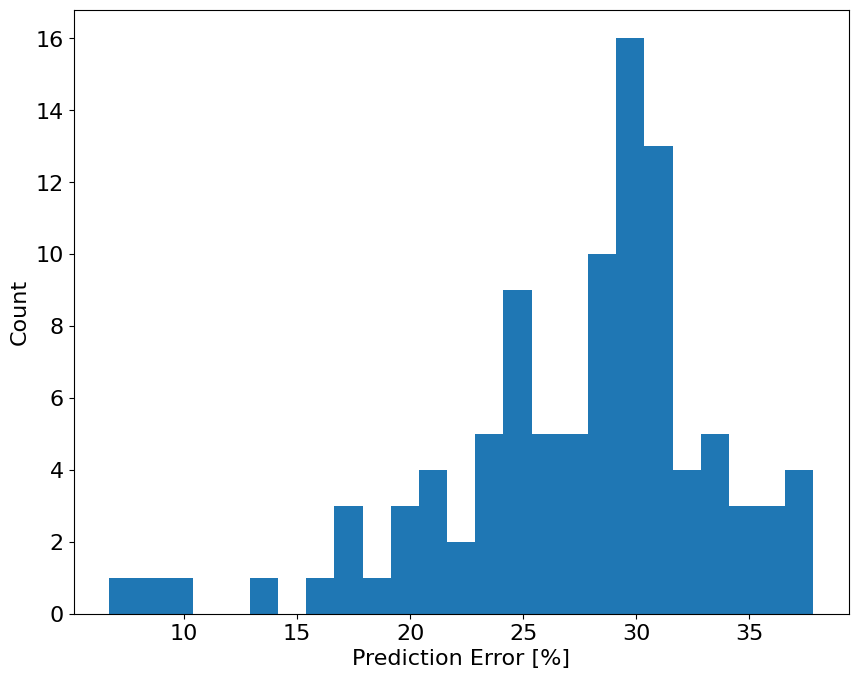}} &
\subfloat[Predictions of the validation data using \ac{RBF} interpolant]{\includegraphics[width=5.0cm]{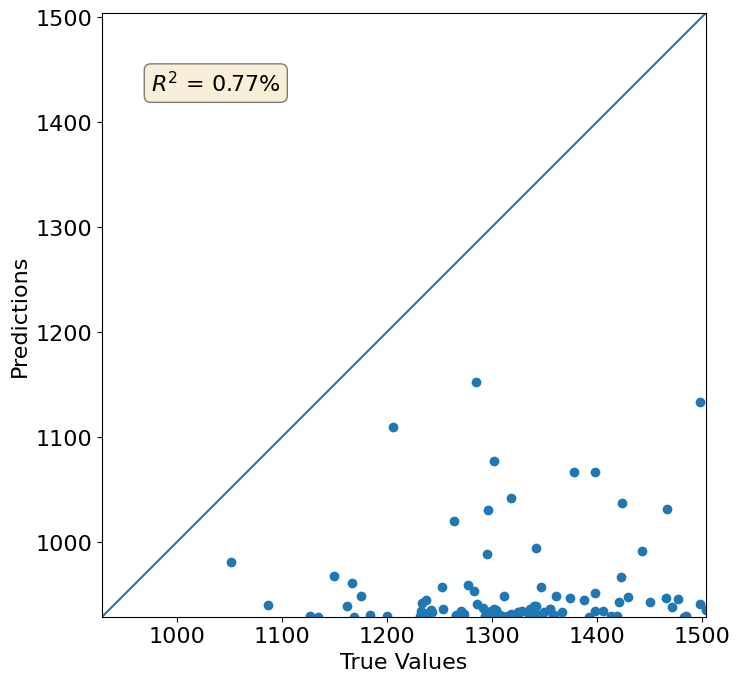}} \\
\subfloat[Histogram of the prediction error using weighted interpolant]{\includegraphics[width=6.0cm]{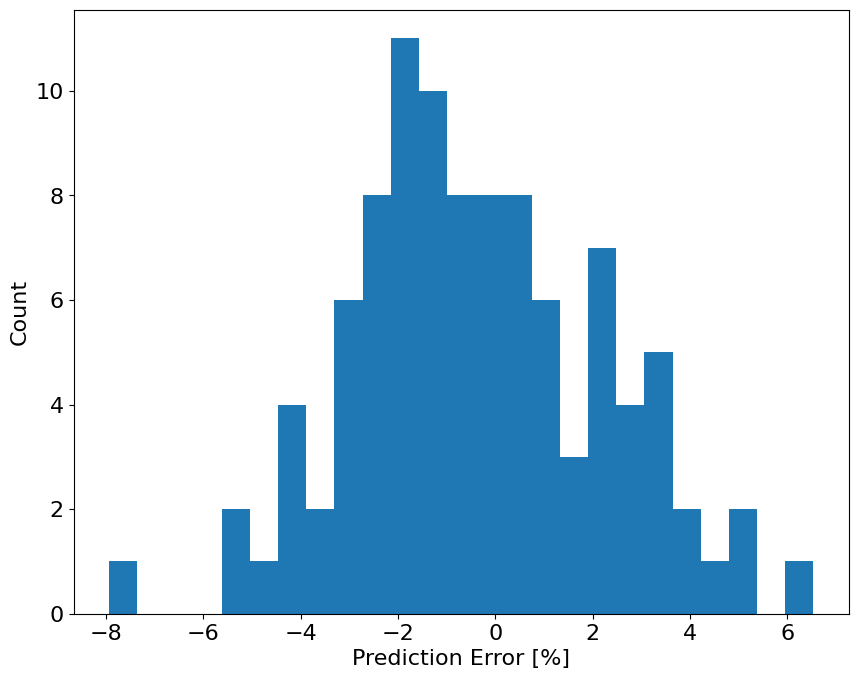}} &
\subfloat[Predictions of the validation data using weighted interpolant]{\includegraphics[width=5.0cm]{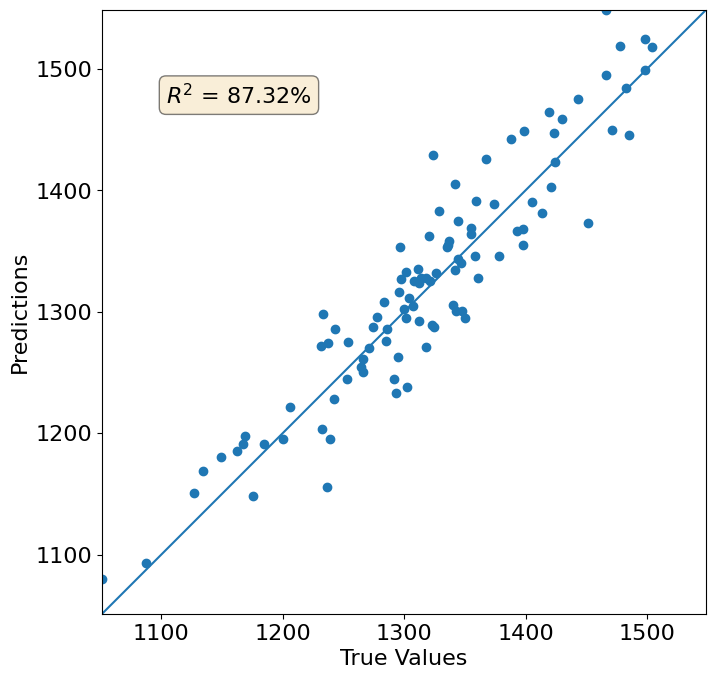}}
\end{tabular}
   \caption{Validation results of nearest neighbor surrogate models (not scaled)}
   \label{fig:results_NN}
\end{figure*}

In linear interpolant (Figure \ref{fig:results_NN} (a) and (b)), hyperplanes created between the closest inputs. \ac{RBF} interpolant  takes the form of a weighed sum of radial basis functions. \ac{RBF} interpolant requires parameter tuning. In Figure \ref{fig:results_NN} (c) and (d), number of neighbors are set to $5$ and \ac{RBF} is set to $11^{th}$ order. In weighed interpolant, weights are calculated automatically based on the distance and distance effect. In nearest neighbor method, the best result is obtained by using weighted interpolant as shown in Figure \ref{fig:results_NN} (e) and (f).

 \bibliographystyle{elsarticle-num} 
 \bibliography{main}





\end{document}